\documentclass[conference]{IEEEtran}
\IEEEoverridecommandlockouts
\usepackage{amsmath,amssymb,amsfonts}
\usepackage{algorithmic}
\usepackage{graphicx}
\usepackage{textcomp}
\usepackage{xcolor}
\usepackage{comment}
\usepackage[utf8]{inputenc}
\usepackage[backend=biber]{biblatex}
\def\BibTeX{{\rm B\kern-.05em{\sc i\kern-.025em b}\kern-.08em
    T\kern-.1667em\lower.7ex\hbox{E}\kern-.125emX}}
\usepackage{titlesec}

\titleformat{\paragraph}[runin]{\normalfont\normalsize\bfseries}{}{0em}{}

\begin{document}

\title{CYGENT: A cybersecurity conversational agent with log summarization powered by GPT-3   \\
%CYGENT: A GPT-3 Powered Conversational Agent for Enhanced Cybersecurity Assistance and Intelligent Log Analysis

\thanks{This work has been funded by the European Commission grants IDUNN (101021911) and NEUROCLIMA (101137711).}
}

\author{\IEEEauthorblockN{1\textsuperscript{st}  Prasasthy Balasubramanian}
\IEEEauthorblockA{\textit{Center for Ubiquitous Computing} \\
\textit{University of Oulu}\\
prasasthy.balasubramanian@oulu.fi}
\and
\IEEEauthorblockN{2\textsuperscript{nd} Justin Seby}
\IEEEauthorblockA{\textit{Biomimetics and Intelligent Systems Group} \\
\textit{University of Oulu}\\
justin.seby@oulu.fi}
\and
\IEEEauthorblockN{3\textsuperscript{rd} Panos Kostakos}
\IEEEauthorblockA{\textit{Center for Ubiquitous Computing} \\
\textit{University of Oulu}\\
panos.kostakos@oulu.fi}

}

\maketitle

\begin{abstract}
In response to the escalating cyber-attacks in the modern IT and IoT landscape, we developed CYGENT, a conversational agent framework powered by GPT-3.5 turbo model, designed to aid system administrators in ensuring optimal performance and uninterrupted resource availability. This study focuses on fine-tuning GPT-3 models for cybersecurity tasks, including conversational AI and generative AI tailored specifically for cybersecurity operations. CYGENT assists users by providing cybersecurity information, analyzing and summarizing uploaded log files, detecting specific events, and delivering essential instructions. The conversational agent was developed based on the GPT-3.5 turbo model. We fine-tuned and validated summarizer models (GPT3) using manually generated data points. Using this approach, we achieved a BERTscore of over 97\%, indicating GPT-3's enhanced capability in summarizing log files into human-readable formats and providing necessary information to users. Furthermore, we conducted a comparative analysis of GPT-3 models with other Large Language Models (LLMs), including CodeT5-small, CodeT5-base, and CodeT5-base-multi-sum, with the objective of analyzing log analysis techniques. Our analysis consistently demonstrated that Davinci (GPT-3) model outperformed all other LLMs, showcasing higher performance. These findings are crucial for improving human comprehension of logs, particularly in light of the increasing numbers of IoT devices. Additionally, our research suggests that the CodeT5-base-multi-sum model exhibits comparable performance to Davinci to some extent in summarizing logs, indicating its potential as an offline model for this task.
\end{abstract}

\begin{IEEEkeywords}
AI chatbot, Cybersecurity, GPT-3, CodeT5, Conversational AI, Generative AI, Log summarizing.
\end{IEEEkeywords}

\section{Introduction}
The emergence of Large Language Models (LLMs), such as GPT (Generative Pre-trained Transformer), has revolutionized AI applications, enabling advancements in text generation, classification, and chatbot functionalities. These innovations have transformed user interfaces (UI) and user experiences (UX), introducing Conversational UIs, Context-aware Assistance, and Personalized UX.

In the cybersecurity domain, leveraging LLMs presents opportunities to enhance threat detection, threat mitigation, and overall security posture. This paper introduces a framework that integrates GPT with conversational user interfaces, facilitating interactions between security analysts and systems. Additionally, the framework incorporates text generation to produce user-friendly summaries of log files. By automating tasks such as log file analysis and summarization, the framework aims to improve efficiency and reduce manual effort in handling large volumes of log data.

The primary contributions of this article can be summarized as follows:
 We introduces a novel framework for a conversational agent powered by GPT models, with the specific goal of effectively assisting security analysts and enhancing the process of log analysis. This framework explores the fine-tuning of GPT-3 models to generate human-readable summaries from complex log data, thereby improving human comprehension of logs, particularly in the context of the increasing number of IoT devices. Additionally, the study includes a comparative assessment to determine whether GPT-3 models excel beyond other prominent LLMs in the task of summarizing log data. Furthermore, the research demonstrates the efficient implementation of the framework, showcasing its ability to transform complex log data into concise summaries and offer valuable support to security analysts.

\section{Related Work}
This section provides a concise overview of recent advancements in chatbot technologies and log analysis techniques within cybersecurity, highlighting their evolution and current limitations.

SecBot \cite{secbot} is an AI-powered tool  that extracts information from conversations, identifies cyberattacks, and offers tailored solutions based on intent classification and entity extraction techniques using NLP and neural networks. It has been evaluated with Rasa 2.0. Similarly, \cite{Patent} developed a chatbot interface for network security software applications, utilizing named entity recognition (NER) and intent classification to understand user intentions and perform network security tasks automatically, although restricted by predefined vocabulary and requiring regular updates.The paper \cite{9530938jj} addressed cyber security risk analysis for Government-to-Citizen (G2C) e-services, focusing on virtual assistants and utilizing the Factor Analysis of Information Risk (FAIR) model for quantitative risk assessment. Their approach involves gathering insights from cybersecurity professionals, and they determine the semantic similarity between cybersecurity terms and threat vectors obtained from expert interviews. However, the method is constrained by predefined vectors stored in the database. The paper \cite{mardini2017messenger} describes the implementation of a bot system connecting Facebook Messenger with IoT devices in a highly secure architecture, aiming to provide developers with a simple, secure, and fast framework for easily integrating the two platforms. One of the latest paper, \cite{li2023chatiot} introduces ChatIoT, which utilizes Large Language Models (LLMs) to process natural language in chat interactions. It focuses on enabling the zero-code generation of Trigger-Action Programs (TAPs), a type of Internet of Things (IoT) application commonly used in smart homes for managing existing devices. 

The paper \cite{JarticleLogassist} presented LogAssist, a novel log summarization approach aiding practitioners in log analysis by organizing logs into event sequences (workflows) and employing n-gram modeling to compress log events. LogAssist reduces log events requiring investigation by up to 99\% and shortens log analysis time by 40\%, according to a user study with 19 participants. While positively received, LogAssist's reliance on predefined hyperparameters suggests avenues for future refinement and adaptation to enhance log analysis practices. In \cite{nimbalkar2016semantic} proposed an approach for log file analysis involving schema analysis and RDF (Resource Description Framework) content sharing, achieved through log content normalization using regular expressions and dictionary-based classifiers. However, the output lacks human readability and may require decoding, and the method's reliance on semantic similarity overlooks contextual information, limiting its effectiveness. The study \cite{ritchey2019naive} proposed log file reduction using the Naive Bayes algorithm, highlighting its effectiveness in feature extraction, although demanding rigorous feature engineering.In paper GPT-2C\cite{setianto2021gpt}, they enhances Intrusion Detection Systems by fine-tuning the GPT-2 model to parse dynamic logs with malicious Unix commands from a Cowrie honeypot, achieving 89\% accuracy.Similarly, \cite{li2013automatic} explored feature extraction and transformation using N-gram techniques and TF-IDF, coupled with K-means clustering adaptations, achieving a notable maximum F-score of 0.943. However, both approaches necessitate substantial manual intervention due to intensive feature engineering, leading to increased implementation time. As part of feature development, in our previous paper \cite{balasubramanian2023transformer} we developed an anomaly detection feature for the CYGENT framework which uses fine tuned GPT3 models for detection of anomalies in the log data.

Our review highlights a gap in leveraging Large Language Models (LLMs) within conversational agents and log file analysis and summarization, despite their extensive use in cybersecurity challenges like Intrusion Detection Systems, Honeypot Log analysis and Distributed Denial of Service (DDoS). Expanding LLM applications in these areas presents an opportunity to achieve SOTA performance. This study aims to explore and exploit these possibilities for further research advancement.

\section{Methodology}
\subsection{Chatbot- Conversational Agent}
To evaluate conversational systems across varied communicative scenarios in the cybersecurity domain, we designed a chatbot, utilizing SOTA algorithms for language comprehension, including the GPT-3.5 \cite{radford2018improving} turbo model, that can handle colloquial dialogues and complex cybersecurity inquiries, offering humanlike responses. The user interface enables intuitive interactions, allowing users to inquire, express opinions, and engage in extended dialogues seamlessly. This innovation marks a new era in conversational AI, particularly applicable in cybersecurity, facilitating evaluation across diverse communicative scenarios.

\begin{figure}[ht]
    \centering
    \includegraphics[width=1\columnwidth]{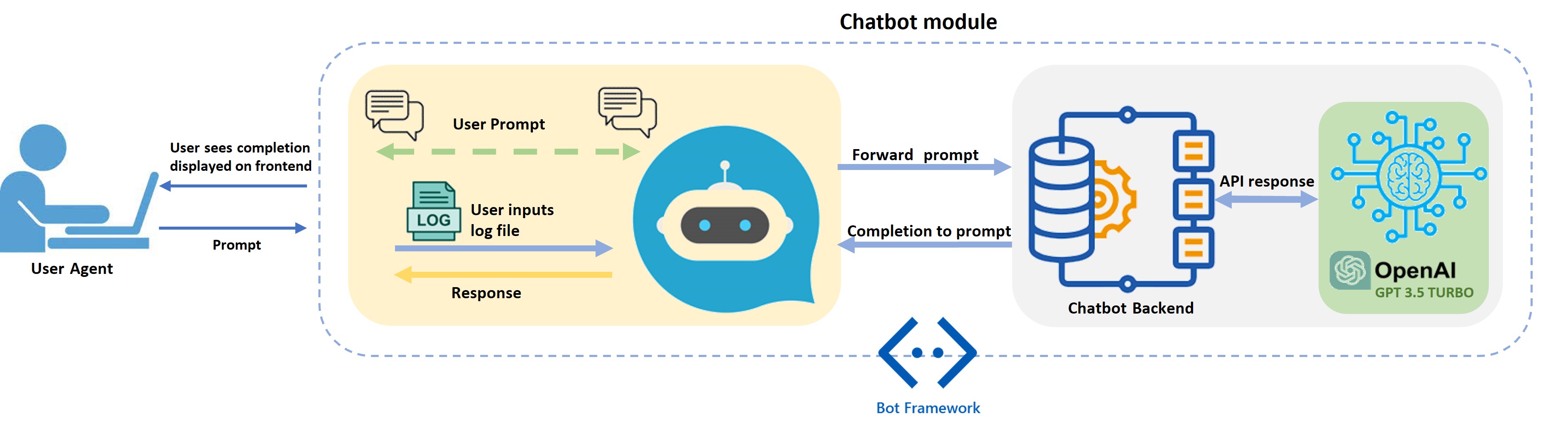} % Adjust the width as needed
    \caption{Chatbot architecture.}
    \label{fig:chatbot_architecture}
\end{figure}

Utilizing GPT-3.5 turbo, we developed a user-friendly interface enabling users to upload log files, especially during incident investigations, and engage in Q\&A sessions with the chatbot for log analysis. The GPT-3.5 turbo processes user inquiries via an API, providing responses regarding data protection, cybersecurity threats, and related details. With a conversation history limit of 4096 tokens, equivalent to approximately 3000 words, the chatbot's short-term memory can manage ongoing interactions, and users can request log file summarization within the chat window. Figure \ref{fig:chatbot_architecture} illustrates the conversational flow and interaction between the user agent and the chatbot.
%Below is the flow chart for chatbot
\subsection{Log Summarizer}
\begin{figure}[ht]
    \centering
    \includegraphics[width=1\columnwidth]{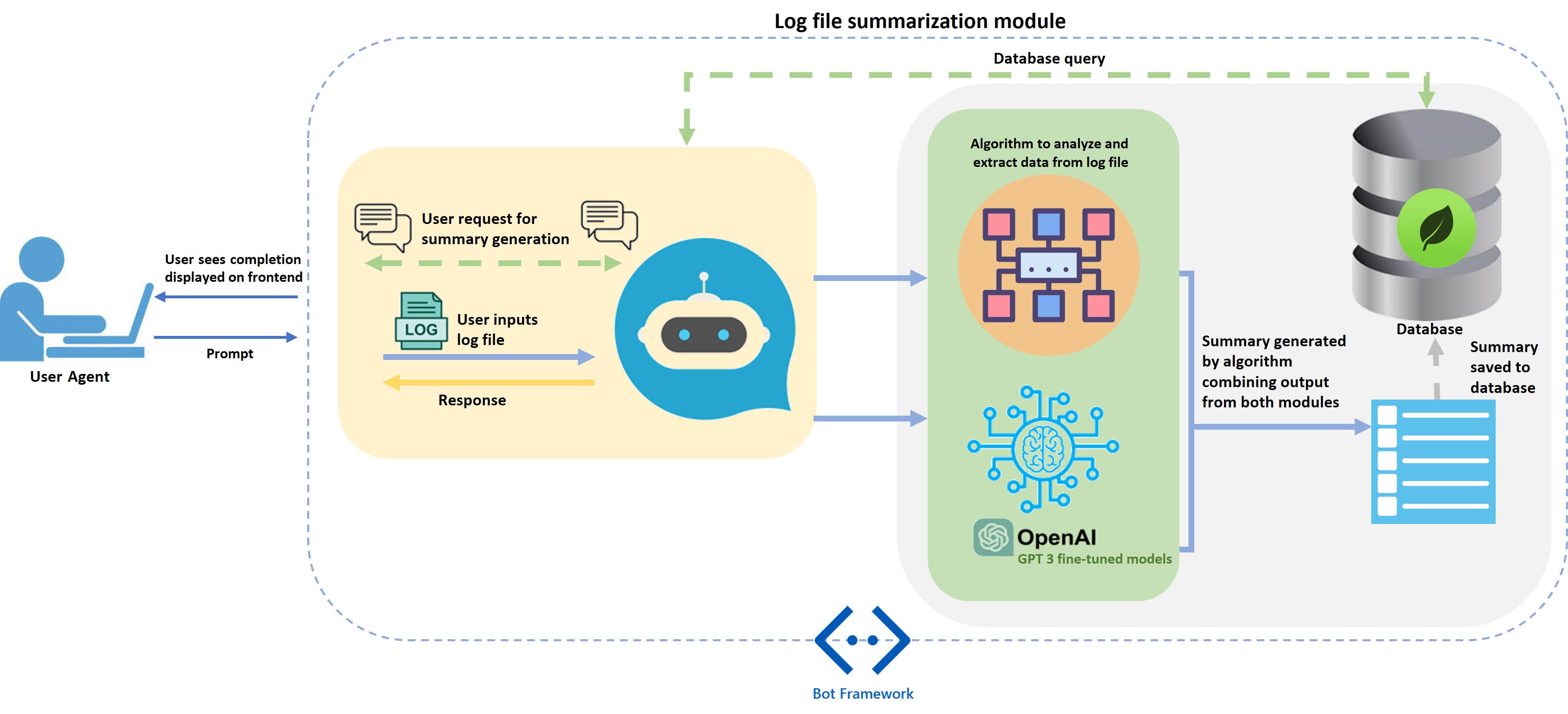} % Adjust the width as needed
    \caption{Summarization architecture.}
    \label{fig:summarization_architecture}
\end{figure}

Logs containing massive amounts of information can be computationally challenging. The chatbot's log summarization feature, leveraging GPT-3's language capabilities, is crucial for extracting essential insights from large log datasets, aiding in system behaviour analysis, health monitoring, and issue troubleshooting. Users can upload log files via the user interface, initiating summarization through an API call to GPT-3 models. The resulting human-readable summary facilitates comprehension of complex log data in a user-friendly manner. The system employs a multi-layered approach, combining data-driven and rule-based techniques, to enhance efficiency in summarization. Utilizing MongoDB, a NoSQL database known for flexibility and scalability, the framework stores generated summaries and file contents in a document-oriented manner. This architecture is depicted in Figure \ref{fig:summarization_architecture}, showcasing the flow of data.

First, our summarizer streamlines conversations by simplifying them to their core points, scanning each log line to create a coherent summary that links contextual information. Additionally, our rule-based logic guarantees comprehensive coverage of conversation context by extracting specific events with errors, warnings, exceptions, and other information like event Types, IP addresses, HTTP Status codes, URLs, file\_paths if any, etc. By combining data-driven summarization with rule-based reasoning, our system produces thorough and refined responses.
The training step involves creating a dataset derived from a rule-based logic framework, and we have manually modified some of the data points to create the desired responses. We explored sampled data available from multiple resources \cite{eliasdabbas-weblogs-dataset, jivoi-awesome-ml-cybersecurity} and generated some synthetic using the code provided in \cite{vishnu0399-server-logs-dataset}. We used regular expressions to construct summaries required for the training dataset that encapsulate similar patterns and important information frequently present in the logs. Once the dataset is created, the model proceeds with training. We used 101 data points for fine-tuning Ada, Babbage, Curie and Davinci models available for GPT-3 from OpenAI. As described in \cite{sawicki2023power}, vanilla versions of these models can be fine-tuned with sizes: Ada 2.7B, Babbage 6.7B, Curie 13B, Davinci 175B. We have used 81 data points for the training cycle and 21 data points for the validation cycle. The created validation file had the same format as the training file, and they were mutually exclusive. Evaluation metrics were calculated against validation data at periodic intervals during training.

As part of our comparative analysis, we examined the performance of GPT3 models in comparison to various encoder-decoder CodeT5 models, as detailed in \cite{wang2021codet5}. CodeT5 models are characterized by an encoder-decoder framework, sharing the same architecture as T5, whereas GPT3 models feature a decoder architecture. CodeT5 models demonstrate proficiency in tasks such as code generation, code translation, code defect detection, and code summarization. Since log information follows a syntax comprising alphanumeric characters and natural language, it can be considered as code, justifying the use of CodeT5 models for log data summarization. We fine-tuned three different versions of CodeT5: small(60M), base (220M), and base-multi-sum.

%\begin{table}[h]
%\caption{Data split up for summary generation models}
%\label{tab:Data_sources_used}
%\centering

%\begin{tabular}{ llllll p{0.4cm} } 
%\hline\noalign{\smallskip}
%Model & Train & validation & Epochs & cost & Time consumed\\
%Ada & 81 & 20 & 4 & \$0.03 &  39 mins and 24 secs \\
%Babbage & 81 & 20 & 4 & \$0.05 &  32 mins and 0 secs  \\
%Curie & 81 & 20 & 4 & \$0.26 &  23 mins and 42 secs\\
%Davinci & 81 & 20 & 4  & \$2.56 &  15 mins and 20 secs\\
%\end{tabular}

%\end{table}
%Below is the flow chart for summarizer
\subsection{History tab and feedback data collection}
This module displays the existing details generated and identified for an already uploaded file for a particular session. The user can also trigger data generation if it was not previously generated. It also has the feature to regenerate the results again for each file and corresponding tasks. Another important feature is to collect data or feedback from users and save it in a database, which can be further utilized for retraining the model. This helps to refine the existing models with annotated data without manual intervention from the user. The framework is constructed entirely in Python, encompassing both the user interface and back-end functionalities. Key technical components include Gradio (3.36.1) and gradio-client (0.2.9), open-source Python libraries facilitating the development of user interfaces and widely used for deploying machine learning models. Gradio is easily customizable and integrates seamlessly with popular Python libraries such as Scikit-learn, PyTorch, NumPy, seaborn, pandas, and TensorFlow. OpenAI (0.27.8) is utilized for accessing the API provided by OpenAI, enabling fine-tuning and interaction with GPT-3 Large Language Models. We use Pymongo (4.4.1) to store data from uploaded log files and model-generated summaries. This allows us to reproduce responses and maintain a history of files and corresponding data for the session.
  
 \begin{figure}[ht]
    \centering
    \includegraphics[width=1\columnwidth]{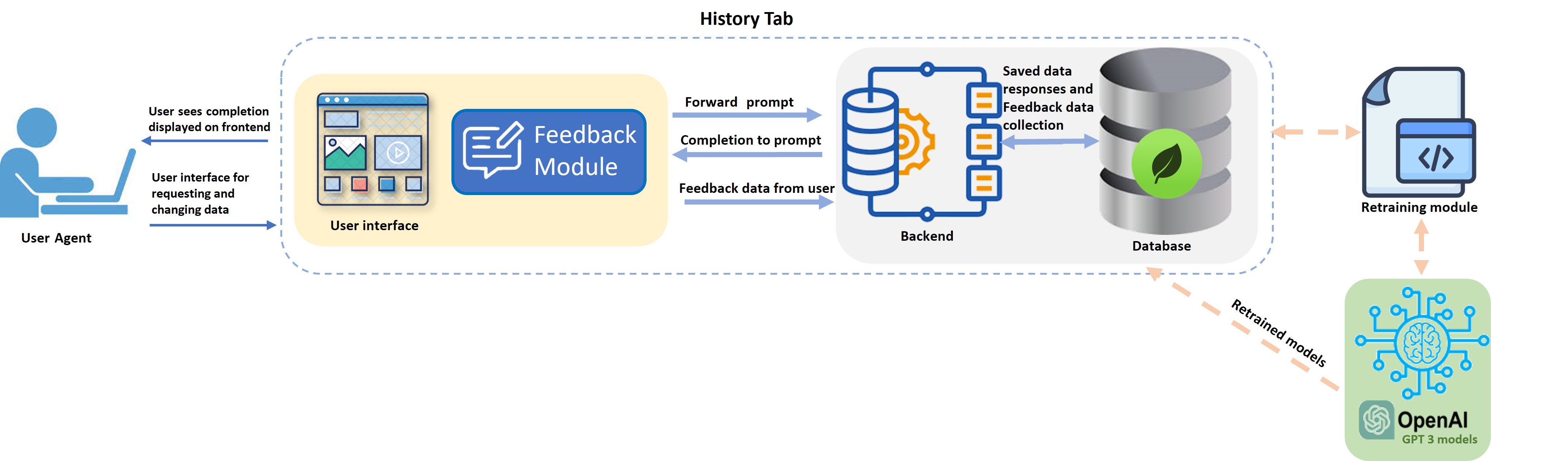} % Adjust the width as needed
    \caption{History data flow architecture.}
    \label{fig:history_architecture}
\end{figure}
%Below is the flow chart for history tab

%Below is the database mapping 
\subsection{Evaluation metrics}
\textbf{OpenAI metrics}: To evaluate the GPT3 fine-tuned models, we utilized Weights and Biases (WandB) platform \cite{wandb}. As the summarization task is generative, key metrics such as training loss, training token accuracy, validation loss, and validation token accuracy were employed. The evaluation parameters included elapsed\_tokens and elapsed\_examples, denoting the total number of tokens and examples seen by the model, respectively. Additionally, metrics like training\_loss, training\_sequence\_accuracy, training\_token\_accuracy, validation\_loss, validation\_sequence\_accuracy, and validation\_token\_accuracy were utilized to assess model performance during training and validation phases for GPT3.

\textbf{Other metrics}: Two other metrics are employed to evaluate the quality of manually generated summaries. Recall-Oriented Understudy for Gisting Evaluation (ROUGE) assesses the word overlap between the generated output and the reference text, with various versions targeting different lengths of word or n-gram overlap. Specifically, ROUGE-L measures the longest common subsequence between the two texts. ROUGE-N computes the ratio of overlapping n-grams to the total number of n-grams in the reference text. Also, the BERTScore \cite{zhang2019bertscore} computes embeddings for words or subwords in both texts and then measures similarity. This approach provides a more nuanced comparison beyond exact word or sequence matches.

%\begin{equation}

%\begin{equation}
%\text{BERTScore} = \text{Cosine Similarity between embeddings}
%\end{equation}

\section{Implementation and Results}
\subsection{Chatbot UI}
\begin{figure} [ht]
    \centering
    \includegraphics[width=1\columnwidth]{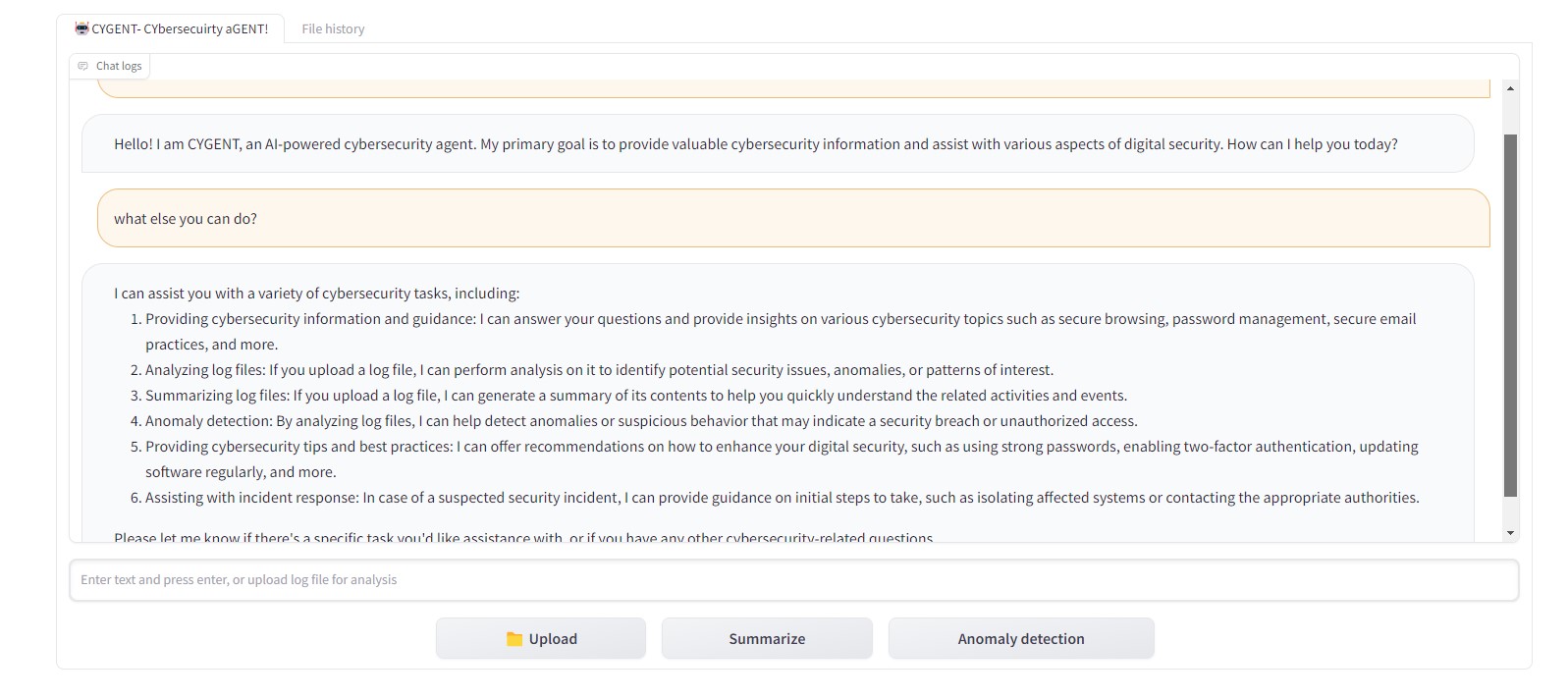}
    \caption{Chatbot User Interface.}
    \label{chatlog}
\end{figure}

The chatbot UI features two main tabs: one for uploading log files, one for summarization and one for displaying chat logs. Another is where users can review previous uploads and their summaries. Chat logs are limited to 4096 tokens (around 3000 words). Refer to Figure \ref{chatlog} for a snapshot of the chat log tab, listing its capabilities to the user.

\subsection{Summarizer}
The results were collected and visualized using plots in WandB for GPT3 models. Analysis was done for details of the token and example counts observed by each model at two key stages, 41 (middle step) and 81 (final step), in midway and at conclusion of training/validation.

\textbf{WandB evaluation}: During the training cycle, Ada exhibited improvement, with training loss decreasing from around 0.137 to 0.0958, indicating closer alignment between model predictions and actual values. Training sequence accuracy and token accuracy also increased, suggesting enhanced capture of complex patterns in the training data. In the validation cycle, there was a decline in validation sequence accuracy, although validation token accuracy notably increased. Babbage showed improvement in the training cycle, with training loss decreasing from approximately 0.165 to 0.124. However, there was a slight decrease in token accuracy from 92.22\% at Step 41 to 91.58\% at Step 81. In the validation phase, there was a noticeable decline in validation sequence accuracy, along with a decrease in validation token accuracy. Curie demonstrated improved convergence during the training cycle, with stable training sequence accuracy and a slight increase in token accuracy. However, validation token accuracy exhibited a decline, while validation sequence accuracy improved between steps. Davinci showed impressive progress during the training cycle, with significant decreases in training loss and increases in training sequence accuracy and token accuracy. However, there was a decline in the accuracy of validation tokens and sequences.

\textbf{Manual evaluation for comparative analysis and model selection}: Our manual evaluation involved generating summaries for five data points using all fine-tuned models. Subsequently, we assessed these summaries using evaluation metrics ROUGE-N, ROUGE-L, and BERTScore. The results are detailed in Tables \ref{tab:rouge-1_F1Score_output_eval}, \ref{tab:rouge-1_F2Score_output_eval}, \ref{tab:rouge-1_FlScore_output_eval}, and \ref{tab:BERTFScore_output_eval}. 

Analyzing the ROUGE-1 (F-Score) presented in Table \ref{tab:rouge-1_F1Score_output_eval}, Davinci consistently achieves high scores, while Ada trails behind the other models. CodeT5-base-multi-sum performs better than all GPT3 models (Ada, Curie, and Babbage) other than Davinci, which shows that CodeT5 models are a good choice for the log summarization task. Regarding the ROUGE-2 (F-Score) in Table \ref{tab:rouge-1_F2Score_output_eval}, Davinci and CodeT5-base-multi-sum appear to perform relatively well across multiple prompts. If considering the average, CodeT5-base-multi-sum sometimes outperforms Davinci in certain instances. Davinci and CodeT5-base-multi-sum appear to perform relatively well across multiple prompts based on the ROUGE-L F-scores shown in Table \ref{tab:rouge-1_FlScore_output_eval}. Lastly, examining the BERTScore (F-Score) outlined in Table \ref{tab:BERTFScore_output_eval} for all prompts, Davinci consistently achieves the highest BERTScore F-scores among all models, indicating that it has the highest token-level similarity with the reference summaries. The performance of other models varies across prompts, but generally, CodeT5-base-multi-sum and Curie also achieve high BERTScore F-scores across prompts. Ada, although still performing reasonably well, tends to have slightly lower BERTScore F-scores compared to other models, especially Davinci. Overall, the analysis suggests that Davinci and CodeT5-base-multi-sum perform well and are eligible options for the final implementation.

\begin{table*}[hbt!]
\caption{ROUGE-1-1(F-score) for summaries}
\label{tab:rouge-1_F1Score_output_eval}
\centering

\begin{tabular}{p{0.9in} p{0.5in}p{0.5in}p{0.5in}p{0.5in}p{0.9in}p{1.2in}p{0.9in}}
\hline\noalign{\smallskip}
Prompt number & Ada & Curie & Babbage & Davinci & CodeT5-small & CodeT5-base-multi-sum & CodeT5-base\\
\noalign{\smallskip}\hline\noalign{\smallskip}
Prompt-1 &  0.088888 & 0.271604 & 0.261904  & 0.532258 & 0.394366 & 0.459770 & 0.531645\\
Prompt-2 &  0.654205 & 0.769230 & 0.579999  & 0.819999 & 0.472222 & 0.824742 & 0.589743\\
Prompt-3 &  0.548387 & 0.564705 & 0.492537  & 0.743362 & 0.568181 & 0.695652 & 0.260869\\
Prompt-4 &  0.366412 & 0.656488 & 0.715447  & 0.560747 & 0.320987 & 0.660194 & 0.703703\\
Prompt-5  & 0.491803 & 0.606896 & 0.725663  & 0.749999 & 0.337662 & 0.646464 & 0.543478\\
%Prompt-6  &  0.350000 & 0.268293 & 0.537634  & 0.505263 \\
%Prompt-7  &  0.555556 & 0.622517 & 0.649573  & 0.619048 \\
\end{tabular}
\end{table*}

\begin{table*}[hbt!]
\caption{ROUGE-2(F-score) for summaries}
\label{tab:rouge-1_F2Score_output_eval}
\centering

\begin{tabular}{p{0.9in} p{0.5in}p{0.5in}p{0.5in}p{0.5in}p{0.9in}p{1.2in}p{0.9in}}
\hline\noalign{\smallskip}
Prompt number & Ada & Curie & Babbage & Davinci & CodeT5-small & CodeT5-base-multi-sum & CodeT5-base\\
\noalign{\smallskip}\hline\noalign{\smallskip}
Prompt-1 & 0.037037 & 0.10638 & 0.040404 & 0.329113 & 0.305882 & 0.245283 & 0.442105\\
Prompt-2 & 0.439716 & 0.46774 & 0.351999 & 0.617886 & 0.390804 & 0.756302 & 0.473118\\
Prompt-3 & 0.321428 & 0.31372 & 0.317647 & 0.633802 & 0.440366 & 0.563636 & 0.162790\\
Prompt-4 & 0.226415 & 0.39751 & 0.533333 & 0.377952 & 0.288659 & 0.495867 & 0.499999\\
Prompt-5 & 0.298850 & 0.39999 & 0.510344 & 0.538922 & 0.274509 & 0.453124 & 0.327586\\
%Prompt-6  &  0.150943 & 0.060000 & 0.307692  & 0.278689 \\
%Prompt-7  &  0.347826 & 0.395833 & 0.425532  & 0.454545 \\

\end{tabular}
\end{table*}

\begin{table*}[hbt!]
\caption{Rogue-l(F-score) for summaries}
\label{tab:rouge-1_FlScore_output_eval}
\centering

\begin{tabular}{p{0.9in} p{0.5in}p{0.5in}p{0.5in}p{0.5in}p{0.9in}p{1.2in}p{0.9in}}
\hline\noalign{\smallskip}
Prompt number & Ada & Curie & Babbage & Davinci & CodeT5-small & CodeT5-base-multi-sum & CodeT5-base\\
\noalign{\smallskip}\hline\noalign{\smallskip}
Prompt-1 & 0.088888 & 0.246913 & 0.238095 & 0.516129 & 0.338028 & 0.436781 & 0.531645 \\
Prompt-2 & 0.598130 & 0.749999 & 0.559999 & 0.819999 & 0.472222 & 0.824742 & 0.589743 \\
Prompt-3 & 0.516129 & 0.564705 & 0.477611 & 0.725663 & 0.568181 & 0.695652 & 0.260869 \\
Prompt-4 & 0.366412 & 0.656488 & 0.682926 & 0.560747 & 0.320987 & 0.660194 & 0.703703 \\
Prompt-5 & 0.475409 & 0.593103 & 0.725663 & 0.749999 & 0.337662 & 0.646464 & 0.543478 \\
%Prompt-6  &  0.350000 & 0.268293 & 0.516129  & 0.484211 \\
%Prompt-7  &  0.555556 & 0.609272 & 0.649573  & 0.619048 \\

\end{tabular}
\end{table*}
\begin{table*}[hbt!]
\caption{BERTscore(F-score) for summaries}
\label{tab:BERTFScore_output_eval}
\centering

\begin{tabular}{p{0.9in} p{0.5in}p{0.5in}p{0.5in}p{0.5in}p{0.9in}p{1.2in}p{0.9in}}
\hline\noalign{\smallskip}
Prompt number & Ada & Curie & Babbage & Davinci & CodeT5-small & CodeT5-base-multi-sum & CodeT5-base\\
\noalign{\smallskip}\hline\noalign{\smallskip}
Prompt-1 & 0.806917 & 0.865538 & 0.880915 & 0.932380 & 0.898788 & 0.881425 & 0.881425\\
Prompt-2 & 0.942412 & 0.963497 & 0.935355 & 0.958671 & 0.917082 & 0.966063 & 0.966063\\
Prompt-3 & 0.931022 & 0.915863 & 0.925954 & 0.969649 & 0.920796 & 0.945890 & 0.945890\\
Prompt-4 & 0.916147 & 0.937956 & 0.958410 & 0.909761 & 0.883313 & 0.939660 & 0.939660\\
Prompt-5 & 0.898010 & 0.942987 & 0.939919 & 0.950636 & 0.848164 & 0.927545 & 0.927545\\
%Prompt-6  &  0.90466 & 0.87337 & 0.92875  & 0.92352 \\
%Prompt-7  &  0.89356 & 0.94888 & 0.93918  & 0.95186 \\

\end{tabular}
\end{table*}

\begin{figure}[hbt!]
    \centering
    \includegraphics[width=1\columnwidth]{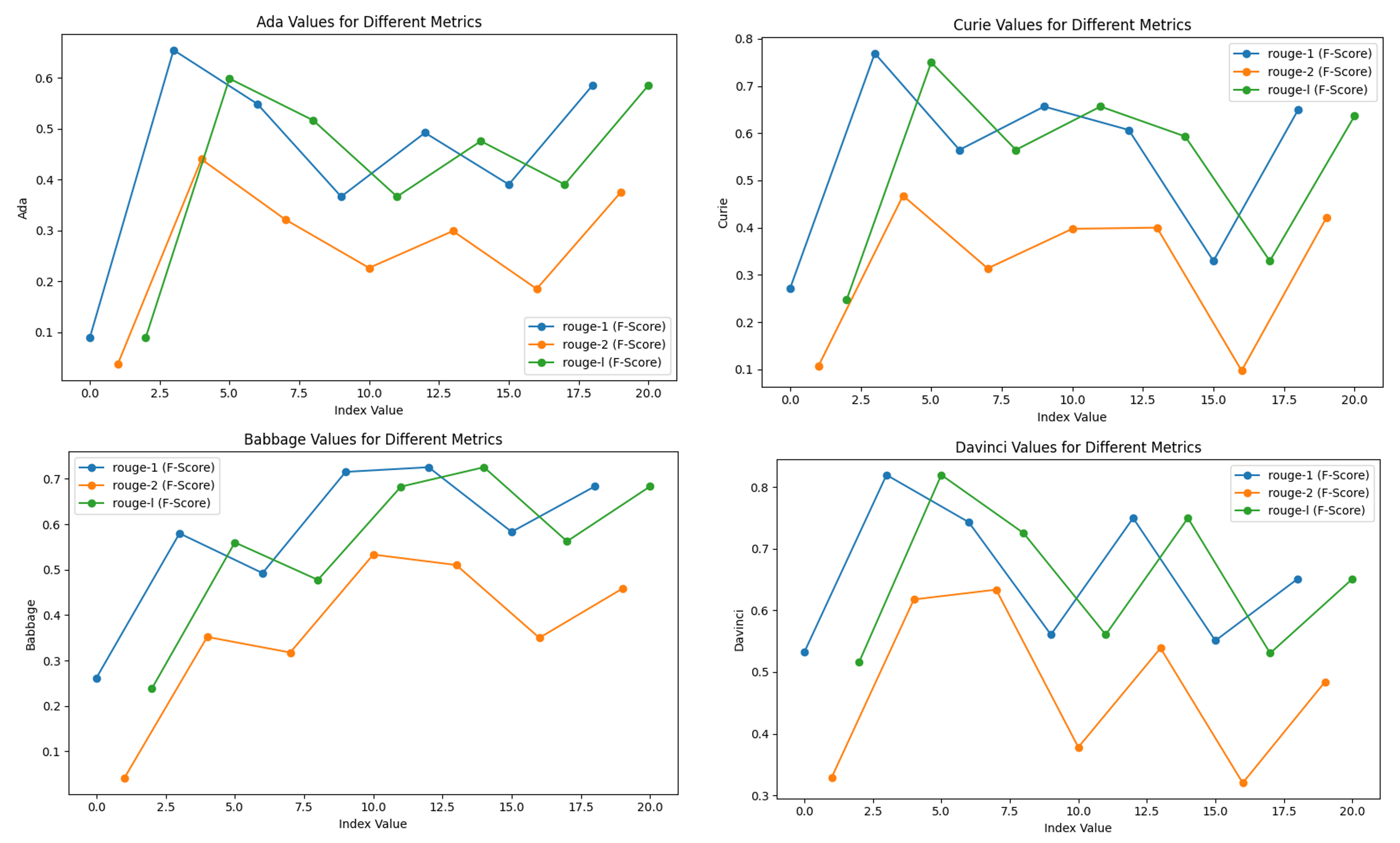}
    \caption{Rogue score for summaries generated by GPT3 models.}
    \label{rogue_model_GPT3}
\end{figure}

\begin{figure}[hbt!]
    \centering
    \includegraphics[width=1\columnwidth]{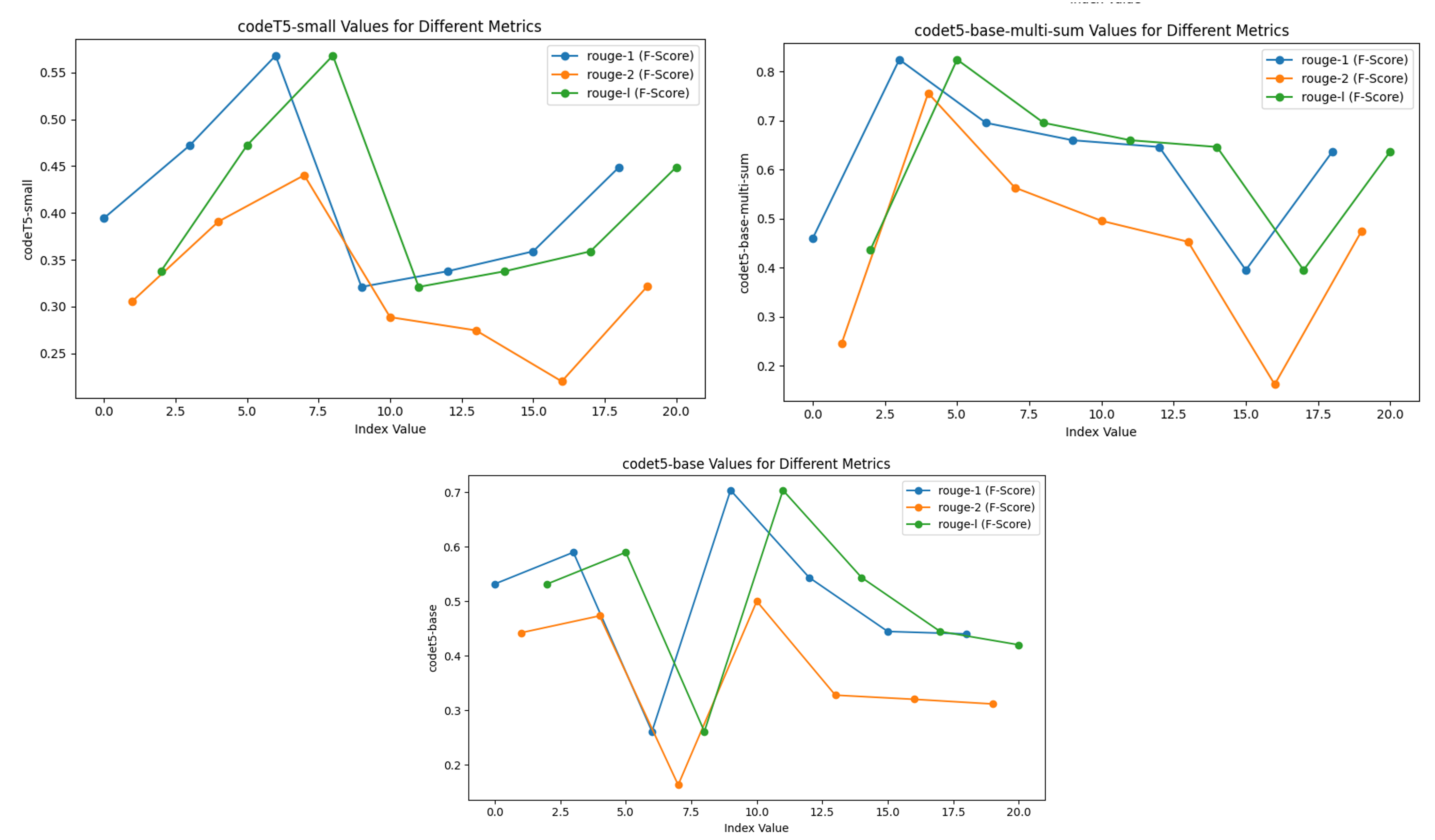}
    \caption{Rogue score for summaries generated by CodeT5 models.}
    \label{rogue_model_codeT5}
\end{figure}

\begin{figure}[hbt!]
    \centering
    \includegraphics[width=1\columnwidth]{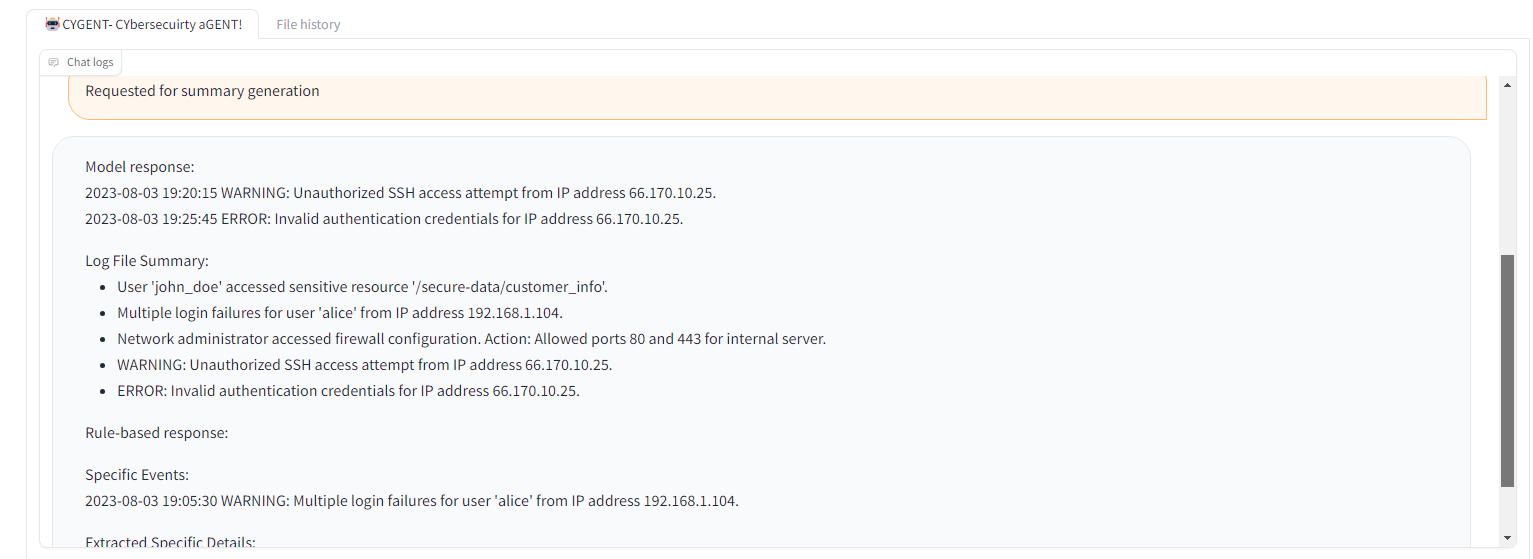}
    \caption{Summary response via chatbot-generated by model}
    \label{chatbot_summ_response}
\end{figure}

Figures \ref{rogue_model_GPT3} and \ref{rogue_model_codeT5} present graphs depicting the executed steps on the X-axis and the corresponding ROUGE scores on the Y-axis. The Davinci model scored higher consistently, while the Ada model scored lower compared to other models at different steps. The CodeT5-base-multi-sum model outperforms Ada and Curie models and moderately outperforms Babbage in most instances. Through our detailed manual evaluation, it was evident that both the Davinci and CodeT5-base-multi-sum models consistently outperformed all other models. Consequently, we integrated the Davinci model as the summarization module within the CYGENT conversation agent framework. An illustration of the model generated user response via the chatbot, after uploading a log file for summary generation, is depicted in Figure \ref{chatbot_summ_response}.
The chatbot offers two distinct response types: one originating from the model and another produced by a rule-based algorithm designed to extract salient details from the uploaded log file. Figure \ref{chatbot_summ_response_rule} below showcases a sample response generated by the rule-based algorithm.

\begin{figure}[hbt!]
    \centering
    \includegraphics[width=1\columnwidth]{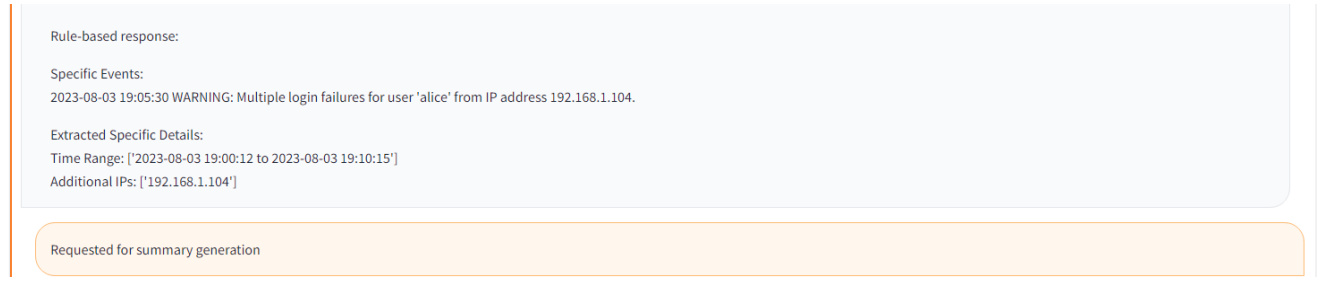}
    \caption{Summary response via chatbot-generated using rule-based algorithm}
    \label{chatbot_summ_response_rule}
\end{figure}

\subsection{History Tab}

\begin{figure}[hbt!]
    \centering
    \includegraphics[width=1\columnwidth]{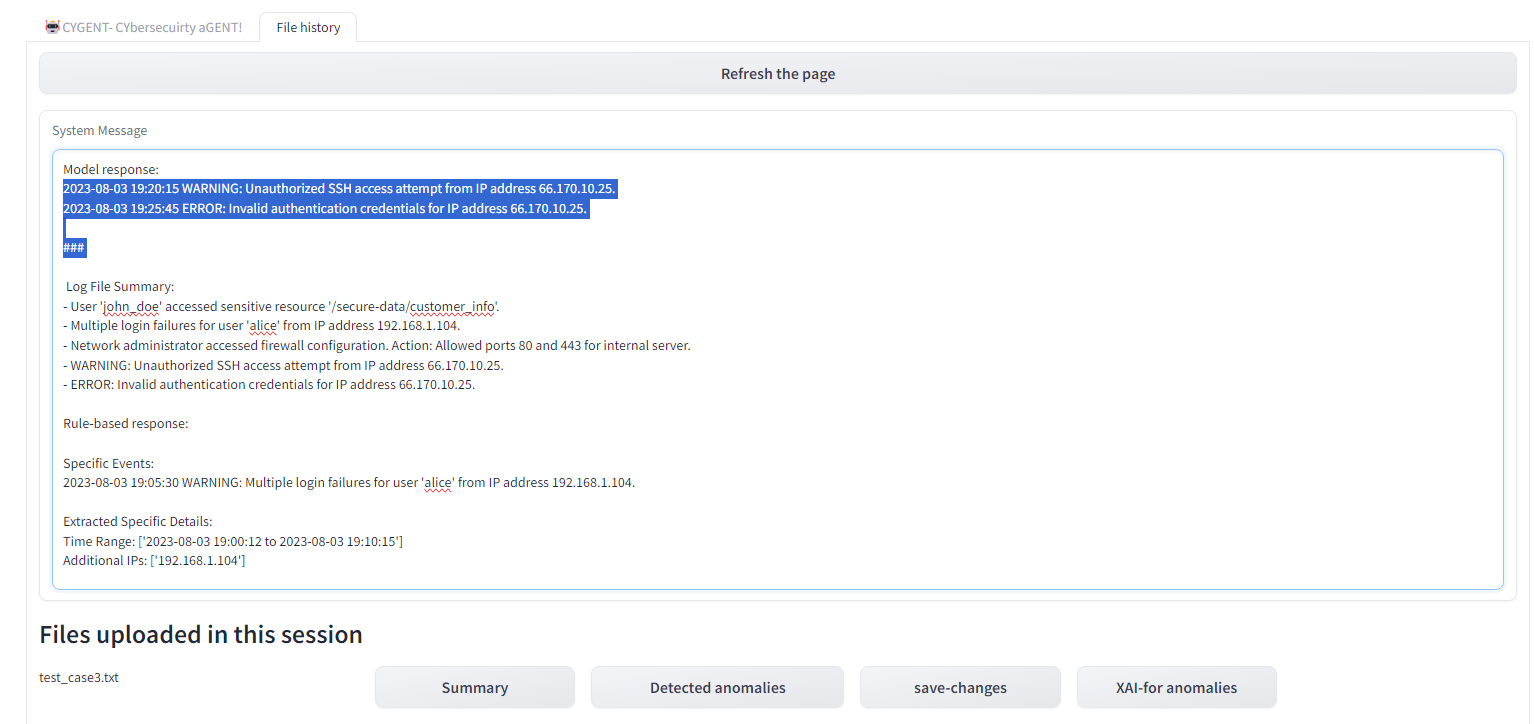}
    \caption{User modifying summary data for changes }
    \label{history_tab_query_change}
\end{figure}

Figure \ref{history_tab_query_change} shows that the users can browse through previously generated summaries in the history tab and modify them for feedback and future reference. This data is stored and utilized for re-training the algorithm to enhance its performance. The systematic process for gathering feedback involves querying the summary, making necessary modifications, and saving changes using the "save-changes" button. Upon successful submission, an acknowledgement response is shown.

%\begin{figure}[hbt!]
%    \centering
%    \includegraphics[width=1\columnwidth]{history_tab.png}
%    \caption{Design of history tab}
%    \label{history_tab}
%\end{figure}

%\begin{figure}[hbt!]
%    \centering
%    \includegraphics[width=1\columnwidth]{history_tab_requested_for_existing_summary.png}
%    \caption{Display of saved summary when queried}
%    \label{history_tab_query}
%\end{figure}
%\begin{figure}[hbt!]
%    \centering
%    \includegraphics[width=1\columnwidth]{response_changes_saved.png}
%    \caption{Response changed successfully }
%    \label{history_tab_query_change_success}
%\end{figure} 

%\item Step4 -Figure \ref{history_tab_query_change_reply} changed response display when queried again.
%\begin{figure}[ht]
%    \centering
%    \includegraphics[width=1\columnwidth]{saving_changed_data.png}
%    \caption{Showing changed data when queried again }
%    \label{history_tab_query_change_reply}
%\end{figure}

\section{Discussion and Future work}
Previous works in cybersecurity chatbots have primarily relied on natural language processing techniques such as entity recognition and intent classification \cite{secbot, Patent}. However, these approaches often depend on predefined keywords or are limited to specific platforms, hindering their widespread adoption \cite{mardini2017messenger, 9530938jj}. In contrast, our methodology leverages Large Language Models (LLMs) like the GPT-3.5 turbo model to achieve superior contextual understanding and platform agnosticism, enabling integration across diverse platforms \cite{radford2018improving}. Our approach utilizes advanced GPT-3 LLM models to bypass time-consuming processes and produce human-readable summaries, even for non-technical users \cite{wandb}. While our fine-tuned models demonstrate progress during training, inconsistencies in validation phases suggest challenges in data generalization. Based on comparative analysis we have found that GPT3 models outperforms other LLMs in the task of summarizing log data. Another significant find is that CodeT5-base-multi-sum show cased a similar performance as GPT3 Davinci model, indicating its eligibility to use as an offline model for the task. 

Our future vision involves expanding and refining our framework to become an essential tool for maintaining secure networks with minimal human intervention. One crucial direction is the integration with Host-based and Network-Based Intrusion Detection Systems (HIDS/NIDS) for continuously monitoring system log files. We aim to transition from manual file uploads to real-time monitoring, enabling seamless production of human-readable summaries as fresh data is recorded, ensuring timely responses to threats or anomalies. Regarding evaluation, We will gather feedback from cybersecurity experts through a survey-based assessment to improve our conversational agent framework for real-world scenarios.

\section{References}
%\bibliographystyle{apacite}
%\bibliography{references}
\printbibliography[heading=none]

\end{document}